\documentclass[aps,pra,twocolumn,showpacs,superscriptaddress,draft,floatfix]{revtex4}
\input{epsf}
\usepackage{amsmath}
\usepackage{amssymb}
\input{epsf}

\newcommand{\rem}[1]{}

\begin{document}

\title{Quantum mechanics in phase space: \\First order comparison between  
the Wigner and the Fermi function}
\author{Giuliano Benenti}
\email{giuliano.benenti@uninsubria.it}
\affiliation{CNISM, CNR-INFM, and Center for Nonlinear and Complex Systems,
Universit\`a degli Studi dell'Insubria, via Valleggio 11, I-22100 Como, Italy}
\affiliation{Istituto Nazionale di Fisica Nucleare, Sezione di Milano,
via Celoria 16, I-20133 Milano, Italy}
\author{Giuliano Strini}
\email{giuliano.strini@mi.infn.it}
\affiliation{Dipartimento di Fisica, Universit\`a degli Studi di Milano,
via Celoria 16, I-20133 Milano, Italy}
\date{\today}

\begin{abstract}
The Fermi $g_F(x,p)$ function provides a phase space description of quantum mechanics
conceptually different from that based on the the Wigner function $W(x,p)$.
In this paper, we show that for a peaked wave packet the $g_F(x,p)=0$ curve
approximately corresponds to a phase space contour level of the Wigner function
and provides a satisfactory description of the wave packet's size and shape.
Our results show that the Fermi function is an interesting tool to 
investigate quantum fluctuations in the semiclassical regime.
\end{abstract}


%3.65.-w Quantum mechanics
%03.65.Sq Semiclassical theories and applications

\maketitle

\section{Introduction}
 
The Wigner phase space representation of 
quantum mechanics~\cite{wigner,wignerreview,zachos} 
is a very useful and enlightening approach. 
It is of practical interest in the description of a broad range  
of physical phenomena, including quantum transport processes in 
quantum optics~\cite{schleich}  
and condensed matter~\cite{jacoboni},
quantum chaos~\cite{saraceno}, 
quantum complexity~\cite{benenti},
decoherence~\cite{zurek},
quantum computation~\cite{saraceno2,terraneo},
and quantum tomography~\cite{wootters}.
Furthermore, the phase space approach brings 
out most clearly the differences and similarities between classical and 
quantum mechanics and offers unique insights into the classical limit
of quantum theory~\cite{heller,berry,brumer,davidovich,habib}.

The Wigner phase space distribution function 
of a quantum state described by a state vector $|\psi\rangle$ reads
\begin{equation}
W(x,p)\equiv \frac{1}{2\pi\hbar}\int_{-\infty}^\infty
dy e^{-\frac{i}{\hbar}py}
\psi\left(x+\frac{y}{2}\right)
\psi^\star\left(x-\frac{y}{2}\right).
\label{wignerfunction}
\end{equation}
(For the sake of simplicity, we consider the case 
of a single particle moving along a straight line).  
The Wigner function provides a pictorial phase-space representation of
the abstract notion of a quantum state and allows us to compute the
quantum mechanical expectation values of observables in terms of 
phase space-averages. 

A different, almost unknown phase space approach is
based on an old paper by Fermi~\cite{fermi}.
As pointed out by Fermi, the state of a quantum system  
may be defined in two completely equivalent ways: 
by its wave function $\psi(x)=\langle x | \psi\rangle$ or 
by measuring a physical quantity $g_F(x,p)$. 
Given the measurement outcome $g_F(x,p)=\bar{g}$,  
$\psi(x)$ is obtained as solution of the eigenvalue equation 
$g_F(x,p)\psi(x)=\bar{g}\psi(x)$, where 
$p=-i\hbar {\partial_x}$.
On the other hand, given the wave function $\psi(x)$ 
it is always possible to find an operator $g_F(x,p)$ such that
\begin{equation}
g_F(x,p) \psi(x)=0.
\label{goperator}
\end{equation}
Using the polar decomposition 
\begin{equation}
\psi(x)=R(x)e^{\frac{i}{\hbar} S(x)},
\end{equation}
where $R(x)$ and $S(x)$ are real functions 
[$R(x) \ge 0$ for any $x$], it is easy to
check that identity (\ref{goperator}) is fulfilled by taking 
\begin{equation}
{g}_F\left(x,-i\hbar \partial_x\right)=
\left[-i\hbar \partial_x -
S'(x)\right]^2+\hbar^2
\frac{R''(x)}{R(x)}.
\label{gfoperator}
\end{equation}
Equation (\ref{goperator}) implies that the corresponding 
physical quantity $g_F(x,p)$ takes the value $\bar{g}=0$. 
The equation
\begin{equation}
g_F(x,p)=
\left[p- S'(x)\right]^2+\hbar^2
\frac{R''(x)}{R(x)}=0
\label{gfimplicit}
\end{equation}
defines a curve in the two-dimensional phase space.
In other words, as expected from Heisenberg uncertainty principle, 
we cannot identify a quantum particle by means of a 
phase-space point $(x,p)$
but we need a curve, $g_F(x,p)=0$. 
Note that it is also possible to write equation (\ref{gfimplicit}) in the form
\begin{equation}
p_\pm= S'(x)\pm
\sqrt{-\hbar^2\frac{R''(x)}{R(x)}}
=m v_M \pm \sqrt{2mV_Q},
\label{ppmFermi}
\end{equation}
where $m$ is the particle mass, $v_M\equiv \frac{1}{m} S'$
the Madelung's velocity ~\cite{madelung}, and 
$V_Q\equiv -\frac{\hbar^2}{2m}\frac{R''}{R}$
the so-called quantum-mechanical
potential~\cite{bohm}.
Equation (\ref{ppmFermi}) locates two points, $(x,p_+)$ and 
$(x,p_-)$, in the phase space for any $x$ such that $R''(x)<0$ and
$R(x)\ne 0$.

The phase space Fermi function $g_F(x,p)$  
and the Wigner function $W(x,p)$ are 
at first sight unrelated.
In particular, for the Fermi function there is 
no interpretation in terms of quasiprobabilities as for 
the Wigner function.
On the other hand, for a Gaussian wave packet the 
$g_F(x,p)=0$ curve is an ellipse
of area $\pi\hbar$~\cite{strini1,strini2} 
and coincides with the phase-space contour level along which 
$W(x,p)=W_{\rm max}/e$, with $W_{\rm max}$ equal to the 
maximum value of $W$.
Different contour levels of $W$ correspond
to different ``equipotential curves'' $g_{F}={\rm constant}$.
The purpose of the present paper is to show that a
similar relation exists when the Gaussian
shape of the wave packet is modified, provided the wave packet remains 
peaked. Finally, we will comment
on the significance of our results in the context of semiclassical
approximations of quantum mechanics.

\section{Gaussian packets}

Let us first consider the Gaussian packet
\begin{equation}
R(x)=G(x)\equiv\frac{1}{\sqrt{\sqrt{\pi}\delta}}
e^{-\frac{(x-x_0)^2}{2\delta^2}},\;\;
S(x)=p_0 x.
\label{eq:gaussian}
\end{equation}
In this case Wigner function (\ref{wignerfunction}) reads
\begin{equation}
W(x,p)=\frac{1}{\pi\hbar} 
e^{-\frac{(x-x_0)^2}{\delta^2}-\frac{\delta^2(p-p_0)^2}{\hbar^2}},
\label{Wgaussian}
\end{equation}
while the Fermi function is given by 
\begin{equation}
g_F(x,p)=\frac{\hbar^2(x-x_0)^2}{\delta^4}+
(p-p_0)^2-\frac{\hbar^2}{\delta^2}.
\label{gFgaussian}
\end{equation}
It is clear from Eqs.~(\ref{Wgaussian}) and (\ref{gFgaussian})
that for Gaussian packet (\ref{eq:gaussian}) we have
\begin{equation}
W(x,p)=\frac{1}{\pi e\hbar }e^{-\frac{\delta^2}{\hbar^2} g_F(x,p)}.
\label{wgf}
\end{equation}
Therefore, 
there is a one to one correspondence between 
the ``equipotential curves'' $g_F(x,p)=K$ and 
$W(x,p)=C$, with $C=\frac{1}{\pi e\hbar}e^{-\frac{\delta^2}{\hbar^2} K}$. 
In particular, the $g_F=0$ curve coincides with the 
curve $W=\frac{1}{\pi e\hbar}=\frac{W_{\rm max}}{e}$,
with $W_{\rm max}=W(x_0,p_0)$ maximum value of $W$.

\section{Non-Gaussian packets}

We now discuss the relation between the Wigner and the Fermi function
when the wave packet is peaked but not Gaussian. 
Assuming a smooth, regular behavior of the packet around its maximum, 
we choose the analytic expression 
$R(x)=C G(x)[1+P(x)]$, with $P(x)$ polynomial [chosen so that $R(x)\ge 0$
for any $x$] and $C$ normalization constant which is irrelevant
for our purposes, while 
$S(x)$ is a polynomial. 
The Wigner function is then given by the Fourier transform 
(\ref{wignerfunction}) of 
\begin{equation}
\begin{array}{c}
\psi\left(x+\frac{y}{2}\right)
\psi^\star\left(x-\frac{y}{2}\right)=
C^2 G\left(x+\frac{y}{2}\right)
G\left(x-\frac{y}{2}\right)
\\
\\
\times \left[1+
P\left(x+\frac{y}{2}\right)+
P\left(x-\frac{y}{2}\right)+
P\left(x+\frac{y}{2}\right) P\left(x-\frac{y}{2}\right)
\right]
\\
\\
\times \, e^{\frac{i}{\hbar}\left[
S\left(x+\frac{y}{2}\right)-
S\left(x-\frac{y}{2}\right)\right]}.
\end{array}
\end{equation}

\subsection{Gaussian $R$}

We computed numerically $W(x,p)$
for several functions $P(x)$, $S(x)$
and found that it strongly depends on $S(x)$, while the dependence
on $P(x)$ is weak, as far as the wave packet remains peaked.
Therefore, we first focus on the case $R(x)=G(x)$. 

\begin{figure}
\centerline{\epsfxsize=8.cm\epsffile{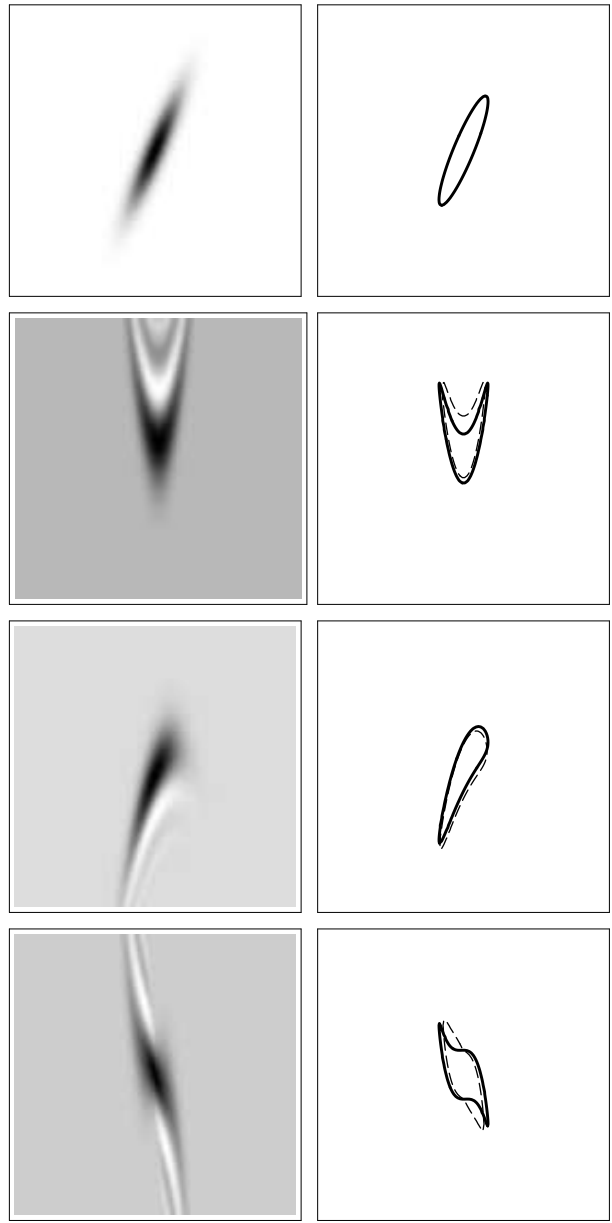}}
\caption{Plots of the Wigner function (left) and of the 
$g_F=0$ curve (right, thick full curves) for $R$ Gaussian and various 
$S$. 
Horizontal axis: $-6\le \tilde{x} \le 6 $, with
$\tilde{x}\equiv \frac{1}{\delta}(x-x_0)$. 
Vertical axis: $-6\le \tilde{p} \le 6 $, with
$\tilde{p}\equiv \frac{\delta}{\hbar}(p-p_0)$. 
From top to bottom: 
$\hbar S =\tilde{x}^2$, 
$\tilde{x}^3$, 
$\tilde{x}^2-\frac{1}{3}\tilde{x}^3$, 
and $-\frac{1}{2}\tilde{x}^4$.
The thin dashed curves in the right plots correspond
to the contour level $W=\frac{W_{\rm max}}{e}$ of the Wigner function.
Note that for the top plot such contour level exactly coincide 
with the $g_F=0$ curve.} 
\label{fig1}
\end{figure}

Wigner functions for $P(x)=0$, namely
\begin{equation}
\psi(x)=G(x)e^{\frac{i}{\hbar}S(x)},
\label{GaussianS}
\end{equation}
and several $S(x)$ are shown in Fig.~\ref{fig1} (left plot)
and compared with the corresponding $g_F=0$ curves (right plots). 
Even though the wave packets in Fig.~\ref{fig1},
with the exception of the top plot [$S(x)\propto x^2$,
corresponding to a squeezed state], 
are far from being Gaussian, the $g_F=0$ curve still 
provides a rather satisfactory description of size and 
shape of the wave packet in phase space. 

This agreement can be explained by the following argument.
If we set $P(x)=0$ and consider the expansion
\begin{equation}
\begin{array}{c}
S\left(x+\frac{y}{2}\right)
-S\left(x-\frac{y}{2}\right)
\\
\\
=S'(x)y+\frac{1}{24}S'''(x)y^3+O(y^5)
\approx S'(x)y,
\end{array}
\end{equation}
we obtain
\begin{equation}
\psi\left(x+\frac{y}{2}\right)
\psi^\star\left(x-\frac{y}{2}\right)
\approx
F(x,y)
e^{\frac{i}{\hbar} S'(x)y},
\label{shift}
\end{equation}
where
$F(x,y)\equiv G\left(x+\frac{y}{2}\right)G\left(x-\frac{y}{2}\right)$.
Therefore, the shift theorem of Fourier transform implies that, if 
\begin{equation}
{\cal F}_y[F(x,y)]=W_G(x,p),
\end{equation}
with ${\cal F}_y$ Fourier transform with respect to the $y$-variable,
then
\begin{equation}
{\cal F}_y\left[F(x,y)e^{\frac{i}{\hbar}S'(x)y}\right]=W_G[x,p-S'(x)].
\end{equation}
We can therefore conclude that the Wigner function corresponding to 
wave vector (\ref{GaussianS}) reads
\begin{equation}
W(x,p)\approx W_G[x,p-S'(x)]
=\frac{1}{\pi\hbar} 
e^{-\frac{(x-x_0)^2}{\delta^2}-\frac{\delta^2 [p-S'(x)]^2}{\hbar^2}}.
\end{equation}
Hence connection (\ref{wgf}) between the Wigner and the Fermi
functions approximately holds around the peak of the wave packet.
The rather good agreement between the $g_F=0$ curve and the contour
level $W=\frac{W_{\rm max}}{e}$ is shown in the right plots of  
Fig.~\ref{fig1}. 
We can conclude that, for peaked packets, the $g_F=0$ curve is close 
to an equipotential curve of the Wigner function, enclosing a phase 
space area of the order of Planck's constant. 

\subsection{Non-Gaussian $R$, $S=0$}

\begin{figure}
\centerline{\epsfxsize=8.cm\epsffile{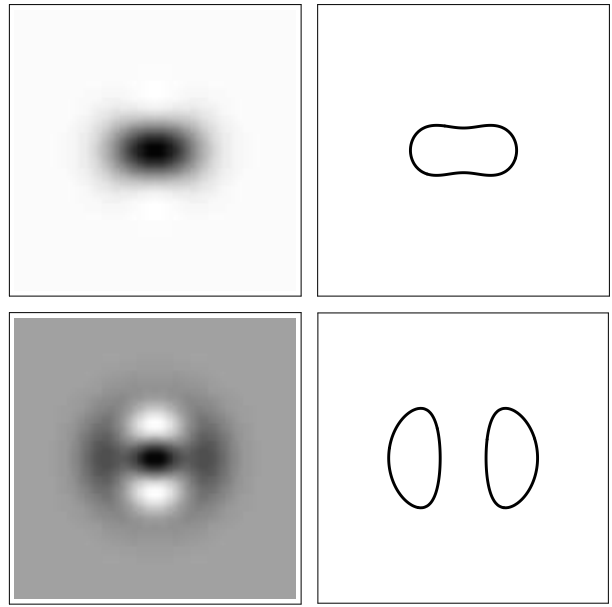}}
\caption{Plots of the Wigner function (left) and of the 
$g_F=0$ curve (right, thick full curves) for 
$P=1+a\tilde{x}^2$, $a=0.3$ (top) and $a=10$ (bottom), 
and $S=0$. 
Horizontal axis: $-4\le \tilde{x} \le 4 $.
Vertical axis: $-4\le \tilde{p} \le 4 $.}
\label{fig2}
\end{figure}

We have seen numerically that the dependence of the Wigner function
on $P(x)$ is weak, as far as the wave packet remains peaked.
As an example, we consider $P(x)=1+a\frac{(x-x_0)^2}{\delta^2}$, with 
$a>0$ so that $R(x)>0$ for any $x$, and $S=0$. For $a<\frac{1}{2}$
the wave function $\psi(x)=R(x)$ has a single maximum at $x=0$
and the $g_F=0$ curve again gives a good representation of 
the phase-space size and shape of the wave packet (see the top 
plots of Fig.~\ref{fig2}). On the other hand, for $a>\frac{1}{2}$
the wave function exhibits two maxima at 
$x_{\pm}=\pm \delta\sqrt{2-\frac{1}{a}}$.
For $a\gg 1$ the two maxima are well separated.
Such a ''cat state'' exhibits non-classical features which impact on  
the structure of the Wigner function (see Fig.~\ref{fig2} bottom left,
for $a=10$). Since $W(x,p=0)$ is the autocorrelation function of
$R(x)$, then it reaches its maximum at $x=0$. On the other hand,
the marginal $\int dp W(x,p)=|\psi(x)|^2=R^2(x)$ exhibits a minimum
at $x=0$ and this is possible thanks to the negative regions of 
$W(x,p)$ (the white regions in Fig.~\ref{fig2} bottom left). 
For this cat state the $g_F=0$ curve captures the two peaks
at $x_\pm$ (see Fig.~\ref{fig2} bottom
right) but not the non-classical phase-space structures of the 
Wigner function. 

\begin{figure}
\centerline{\epsfxsize=8.cm\epsffile{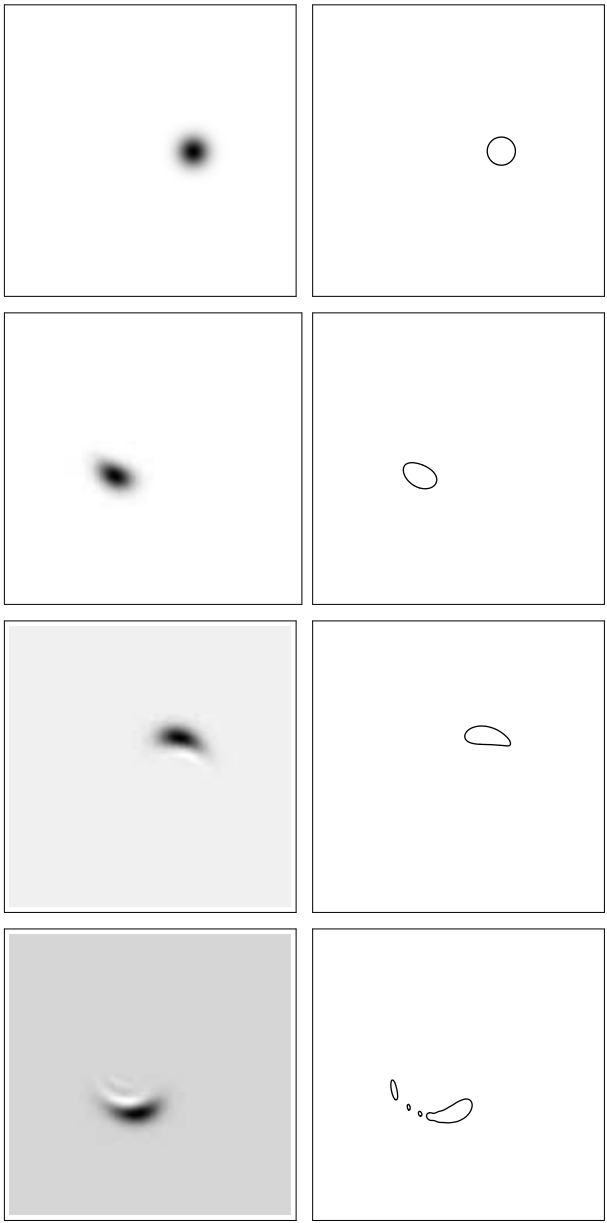}}
\caption{Plots of the Wigner function (left) and of the 
$g_F=0$ curve (right) for the quartic oscillator,
with $\hbar \lambda/\omega=0.01$.
Horizontal axis: $-10\le \tilde{x} \le 10 $, with
$\tilde{x}\equiv \sqrt{\frac{m\omega}{\hbar}}x$.
Vertical axis: $-10\le \tilde{p} \le 10 $, with
$\tilde{p}\equiv \sqrt{\frac{1}{\hbar m \omega}} p$.
From top to bottom: $t/T=0,0.4,0.8,1.2$, with $T=2\pi/\omega$.
The initial Gaussian distribution is centered in 
$(\tilde{x},\tilde{p})=(3,0)$.} 
\label{fig3}
\vspace{-0.4cm}
\end{figure}

\section{A dynamical example: the quartic oscillator}

The $g_F=0$ curve is an interesting tool to investigate quantum fluctuations
in the semiclassical region. As far as the wave packet remains peaked,
that is, before the Ehrenfest time scale~\cite{berman},
its size and phase-space shape can be readily derived from the wave
function (or from a semiclassical approximation of the wave function),
without computing the whole Wigner function. 
Moreover, the distortion of the wave packet 
is directly related to $S'(x)$, that is, to the Madelung's 
velocity. 

As a numerical illustration
of the capability of the $g_F=0$ curve to capture relevant features of
quantum fluctuations, we follow in Fig.~\ref{fig3} the evolution
of the Wigner and Fermi functions for the quartic oscillator Hamiltonian
\begin{equation}
H=\hbar\omega a^\dagger a + \hbar^2 \lambda (a^\dagger)^2 a^2.
\label{eq:quartic}
\end{equation}
Here, $a=\sqrt{\frac{1}{\hbar m \omega}}(m\omega x + i p)$,
$\omega$ is the frequency of the harmonic part of the oscillator
and $\hbar \lambda$ gives the strength of the nonlinearity.  
This model has been widely investigated in the context of quantum
to classical transition~\cite{berman,milburn,angelo,oliveira}
and also used to explain important experimental results~\cite{bloch}.
Model (\ref{eq:quartic}) is integrable, see Refs.~\cite{milburn,oliveira}
for the evolution of classical and quantum phase-space distributions. 
Details on the computation of the Fermi function are given in the 
Appendix. 

In Fig.~\ref{fig3} we compare the evolution the Wigner function with
the evolution of the $g_F=0$ Fermi function, for the quartic oscillator, 
starting from an initial Gaussian wave packet 
$|\alpha\rangle$ ($a|\alpha\rangle=\alpha|\alpha\rangle$). 
It is clear that, as far
as the wave packet remains peaked, the $g_F=0$ curve reproduces both
size and shape of quantum fluctuations. This is the case for times
smaller than the Ehrenfest time scale 
$t_E\sim\frac{1}{\hbar\lambda |\alpha|}$~\cite{berman},
until which the centroid of the wave packet follows a classical 
trajectory. For longer times the Wigner function develops interference
fringes, while the $g_F=0$ function splits into several curves. 
We point out that the $g_F(x,p)=0$ function (\ref{ppmFermi}) singles out 
only two $p$-values (or none) for any $q$. Therefore, it cannot reproduce
the whole phase space structure of the wave packet when the Wigner 
function does not exhibit a single peak but a non-monotonous behavior  
along $p$ [see the bottom plot of Fig.~\ref{fig3} (left)].   

\section{Concluding remarks}

In summary, we have shown that the phase space structure of a peaked
wave packet can be satisfactorily described
by the $g_F=0$ Fermi curve. 
In spite of the fact that the Wigner and the Fermi functions
are at first sight two completely unrelated phase space descriptions of 
quantum mechanics, a link between them 
exists and is based on the shift
theorem of Fourier transform. Such theorem also allows us to 
understand the shape of the Wigner function for perturbed 
Gaussian packets in terms of the Madelung's velocity. 
Our theoretical results, corroborated by numerical simulations 
for the quartic oscillator model, show that the Fermi function is 
an interesting tool to investigate the phase-space size and shape 
of quantum fluctuations in the semiclassical regime. 

While the Fermi function $g_F(x,p)$ fully determines the state of 
a quantum system, the extension of the results obtained in this
paper to generic states encounters difficulties.
Knowledge of the $g_F(x,p)=0$ curve is in general not sufficient.
The complete determination
of the state of a system requires the extension of this curve
to the complex $p$-plane. That is, 
consideration of the complex values of $p_{\pm}$, obtained
from Eq.~(\ref{ppmFermi}) when $R'' >0$, is needed.
We then obtain
$S' = \frac{p_++p_-}{2}$,
$\hbar^2\frac{R''}{R}=
-\left(\frac{p_+-p_-}{2}\right)^2$,
from which the $g_F$ operator (\ref{gfoperator}) and consequently
the wave function $\psi(x)$ are determined.
Any phase space description of quantum mechanics necessarily 
involves features beyond classical intuition: negative quasiprobabilities 
in the case of the Wigner function, complex momenta for the Fermi 
curve. 

\appendix

\section{Fermi function for the quartic oscillator}

We consider as initial condition a Gaussian state
(coherent state for the harmonic part of the oscillator), 
$|\alpha\rangle = \sum_{n=0}^\infty c_n |n\rangle$, 
with $c_n=e^{-\frac{1}{2} |\alpha|^2}\frac{\alpha^n}{\sqrt{n !}}$,
$a|\alpha\rangle = \alpha |\alpha\rangle$, $H|n\rangle = E_n |n\rangle$,
$E_n=\hbar \omega + \hbar^2 \lambda n(n-1)$. The state of the quartic 
oscillator at time $t$ is then given by 
$|\psi(t)\rangle = \sum_{n=0}^\infty c_n e^{-\frac{i}{\hbar} E_n t}|n\rangle$.
In the coordinate representation, 
\begin{equation}
\begin{array}{c}
\phi_n(x)\equiv \langle x | n \rangle 
\\
\\
= 
\left(\frac{m\omega}{\pi\hbar}\right)^{1/4}
\frac{1}{2^{n/2}\sqrt{n!}} 
H_n\left(\sqrt{\frac{m\omega}{\hbar}} x\right)
e^{-\frac{1}{2}\frac{m \omega}{\hbar} x^2},
\end{array}
\end{equation}
where $H_n$ denotes the $n$-th Hermite polynomial. 

In order to compute the Fermi function, we write $\psi=\psi_R+i \psi_I$, 
so that 
\begin{equation}
\begin{array}{c}
\frac{R''}{R}=\frac{1}{(\psi_R^2+\psi_I^2)^2}
(\psi_R \psi_R'+\psi_I \psi_I')^2 
\\
\\
+\frac{1}{\psi_R^2+\psi_I^2}
\left[(\psi_R')^2+(\psi_I')^2+\psi_R \psi_R''+ \psi_I \psi_I''\right],
\end{array}
\end{equation}
\begin{equation}
S'=\frac{\psi_R \psi_I' - \psi_R' \psi_I}{\psi_R^2+\psi_I^2}.
\end{equation}
Finally, the $g_F=0$ curve is obtained from (\ref{gfimplicit}).
Note that 
in computing the derivatives of $\psi_R$ and $\psi_I$, we took advantage
of the relation $H_n'= 2 n H_{n-1}$. This property of Hermite polynomials 
allowed us to avoid numerical errors in the computation of the derivatives
$R''$ and $S'$.

\end{document}